\documentclass[11pt,a4paper]{article}

\usepackage{color,amsmath,amssymb,soul,graphicx} 

\usepackage{geometry}
\usepackage{float}
\usepackage{color}
\usepackage{graphicx}
\usepackage{amssymb}
\usepackage{authblk}
\usepackage{setspace}
\usepackage{soul}
\usepackage[right]{lineno}
\usepackage{cite}

\title{Closed-loop control of a modular neuromorphic biohybrid}

\author[1,2*]{Hanna Keren}
\author[2]{Johannes Partzsch}
\author[1]{Shimon Marom}
\author[2]{Christian Mayr}

\affil[1]{\small{\textit{Network Biology Research Laboratory in Electrical Engineering and Department of Physiology, Biophysics and Systems Biology in Medicine; Technion - Israel Institute of Technology, Israel }}}
\affil[2]{\small{\textit{Institute of Circuits and Systems, Technical University of Dresden, Germany}}}
\affil[*]{\small{Corresponding author: Hanna.keren@nih.gov}}

\begin{document}
\date{}
\maketitle

\begin{abstract}

Neural networks modularity is a major challenge for the development of control circuits of neural activity. Under physiological limitations, the accessible regions for external stimulation are possibly different from the functionally relevant ones, requiring complex indirect control designs. Moreover, control over one region might affect activity of other downstream networks, once sparse connections exist. We address these questions by developing a hybrid device of a cortical culture functionally integrated with a biomimetic hardware neural network. This design enables the study of modular networks controllability, while connectivity is well-defined and key features of cortical networks are accessible. Using a closed-loop control to monitor the activity of the coupled hybrid, we show that both modules are congruently modified, in the macroscopic as well as the microscopic activity levels. Control impacts efficiently the activity on both sides whether the control circuit is an indirect series one, or implemented independently only on one of the modules. Hence, these results present global functional impacts of a local control intervention. Overall, this strategy provides an experimental access to the controllability of neural activity irregularities, when embedded in a modular organization.

\end{abstract}

\section{Author summary}
Using closed-loop neuroscience for stabilizing and modifying irregular brain activities, is becoming increasingly feasible with technological advancement of medical applications. However, at the whole brain level where the relevant neural regions are embedded in a modular organization, such an external intervention can have major functional impacts on other brain regions. Our findings present the full functional integration between a hardware network and a biological culture - a relevant model for neural-machine interfaces. Using this hybrid model, we show that controlling the activity of a single module affects functionally also a sparsely connected downstream region. Hence, an extensive spatial spread of the impacts of closed-loop control to other regions, should be considered.

\newpage
\section{Introduction}

When addressing the question of controllability of neural activity, the inherent modularity of neural networks, in both structure and function, is a major consideration \cite{Mountcastle1997, Rockland1998, erdikiss, Hubelandwiesel, Sporns2000, Derdikman2003, diamond2003, Kumar2010, Pan2010, Boucsein2011}. It has been previously shown that when a modular construct of neural networks is externally stimulated, the stimulation affects also the downstream modules, even when only sparsely connected \cite{Levy2012}. 
In the case of a single neural network, as can be implemented \textit{in-vitro}, an efficient control of activity has been demonstrated and it's impacts could be studied, over a long time scale of many hours (see \cite{Keren2014} and \cite{Feber2010}). Attempts for closing the loop on neural activity \textit{in-vivo} have been also made \cite{Rosin2011b,Bergmann2012,Berenyi2012,Siegle2014}, yet, addressing the question of controllability in a well-defined modular construct, is experimentally challenging.

In order to extend recent studies to the modular level, we combine network control \textit{in-vitro} with biomimetic hardware components (neuromorphic CMOS-based VLSI). These components may be integrated, functionally and physically, with neural systems.
The experimental \textit{in-vitro} setting comprises large-scale, random networks of cortical neurons. It enables long-term stimulation and recordings from tens to hundreds of neurons simultaneously, at high spatio-temporal resolution. This preparation makes key features of \textit{in-vivo} cortical networks accessible, as connectivity, cellular and synaptic physiology, synchronization, learning and most relevant in this context- closed-loop control.

From its inception in 1989 \cite{mead90}, neuromorphic engineering tried to mimic the design and operating principles of neural networks, to develop biomimetic microelectronic devices which implement biological models \cite{bartolozzi07}. So far, the neuromorphic approach has been successful in implementations of sensory functions (e.g. visual processing \cite{koenig02}) and computational functions that rely on building blocks of brain processing (e.g. pattern recognition \cite{qiao15}). From a technical perspective, a variety of circuit techniques have been used to build these functions, from the classic subthreshold approach \cite{mead90} via amplifier/resistor-based replications of behavioral equations \cite{noack10} up to novel semiconductor devices on the nanoscale \cite{mostafa15,du15}.
In this study, a state-of-the-art neuromorphic VLSI chip is designed in the switched-capacitor technique that exhibits finely controllable dynamics with a multiplicity of time-scales and significantly reduces the parameter deviations usually experienced in such systems \cite{henker07,folowosele09b}. The overall system, composed of 9 individual chips and a communication backplane, is able to emulate 2880 neurons (320 per chip) and 12.7M synapses, comparable in size to the recorded region in the neural culture.


We couple this biomimetic hardware network to an \textit{in-vitro} neural culture, and demonstrate the possibility to functionally integrate the two. This is shown by implementing a control circuit, which is modifying the overall activity of the construct by considering only the output activity of the downstream module. In this design, stimulation is applied to the upstream culture (with varying amplitudes), while the activity being monitored is of the downstream hardware module (See the Neural Response Clamp in \cite{Wallach2011,Wallach2013}). This control circuit also represents possible implementations \textit{in-vivo}, where the function monitored is downstream to the stimulated neural network \cite{Walker2012}.  

We also ask whether enforcing different activity levels on one module of the biohybrid, would be indirectly reflected in the other module. This is examined by coupling the hardware network to the activity of a culture which is being independently controlled to different response levels. We show that the downstream hardware module reflects both the macroscopic and microscopic activity features, induced by the controller in the upstream culture. This also suggests that biomimetic networks with sufficiently rich dynamics can serve as a tool for classifying neural network states while overcoming the limitations of sampling through a downstream module. These results are relevant to brain-machine-interface applications, when aiming at communicating with modular neural networks.

\section{Material \& Methods}

\textit{Hardware:} 
A preliminary overview of the NeuroSoC has been published in \cite{noack14}. As the NeuroSoC should only replicate short-term dynamics, it omits the usual synaptic matrix with its long-term plasticity computation \cite{qiao15}. Thus, instead of the conventional neuromorphic chip layout with presynaptic adaptation at the input, a synaptic matrix and postsynaptic neurons, the NeuroSoC can be thought of as a collection of neuron building blocks, the neuron elements (see right half of Figure 1 and respectively Figure 2A for an organizational overview). The neuron elements are grouped in 10 groups of each 32 neuron elements (for a total of 320 neuron elements). Analog biasing voltages for neurons and synapses (e.g. synaptic reversal potentials, membrane resting potentials, etc) are generated via digital-analog converters on chip.
\vspace{+1cm}

\begin{figure}[H]
\vspace{+6em}
\begin{center}
\centerline{\includegraphics[trim= 0cm 10cm 0cm 4cm,width=160mm]{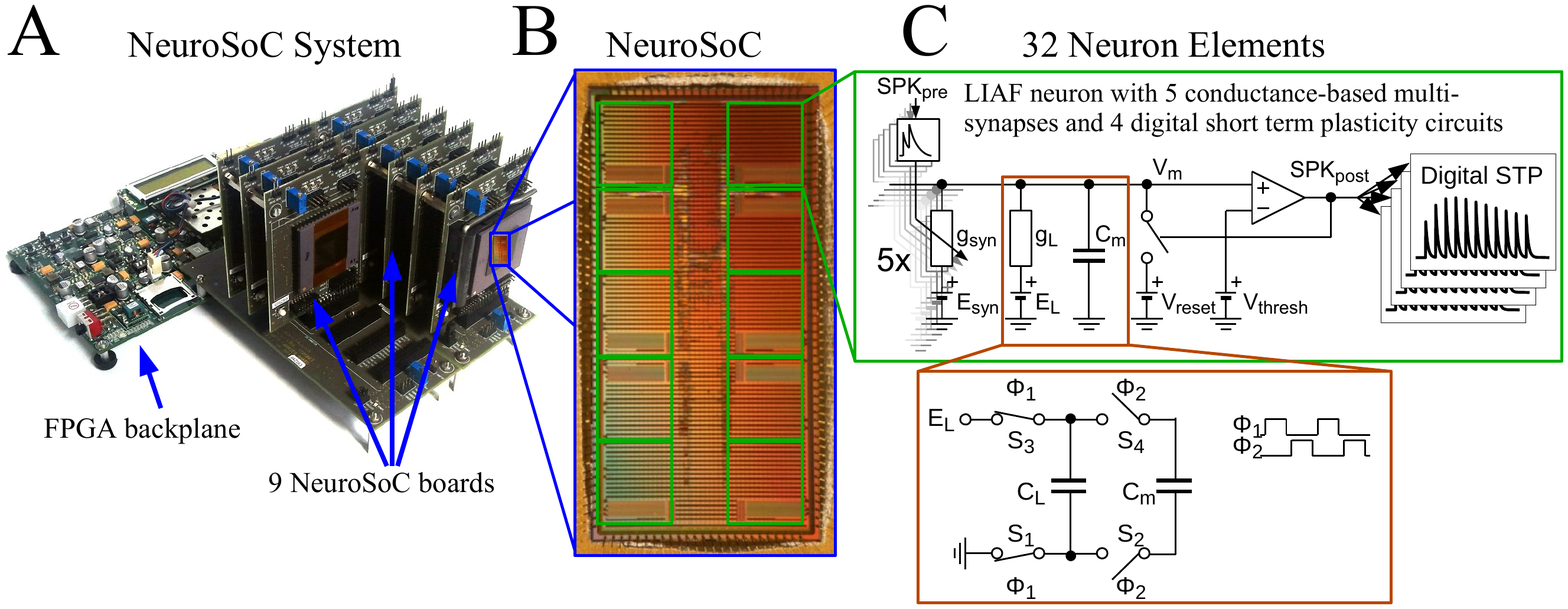}}
\end{center}
\vspace{-0em}
\label{fig_neurosoc}
\end{figure}

\footnotesize{\textbf{Figure 1. The Neuromorphic System-on-Chip (NeuroSoC)} (\textbf{A}) The overall NeuroSoC system, with  9 individual
NeuroSoCs in combined operation; a support system realized on a commercial Field Programmable Gate Array (FPGA) supplies action potential routing and configuration between the individual NeuroSoC chips; Also, the FPGA backplane enables communication to the host PC and the neural culture. (\textbf{B}) The NeuroSoC implements biophysical short term dynamics and a big network size (320 neurons and up to 1.4 million presynaptic spike inputs per chip) for a realistic counterpart to the neural culture. (\textbf{C}) Circuit of one of the 10 neuron groups contained on the NeuroSoC, each with 32 neurons realized in switched capacitor technique (see red insert), each with five types of conductance-based synaptic input.}
\normalsize
\vspace{+1cm}

\begin{figure}[H]
\vspace{-0em}
\begin{center}
\centerline{\includegraphics[trim= 0cm 7cm 0cm 6cm, width=110mm]{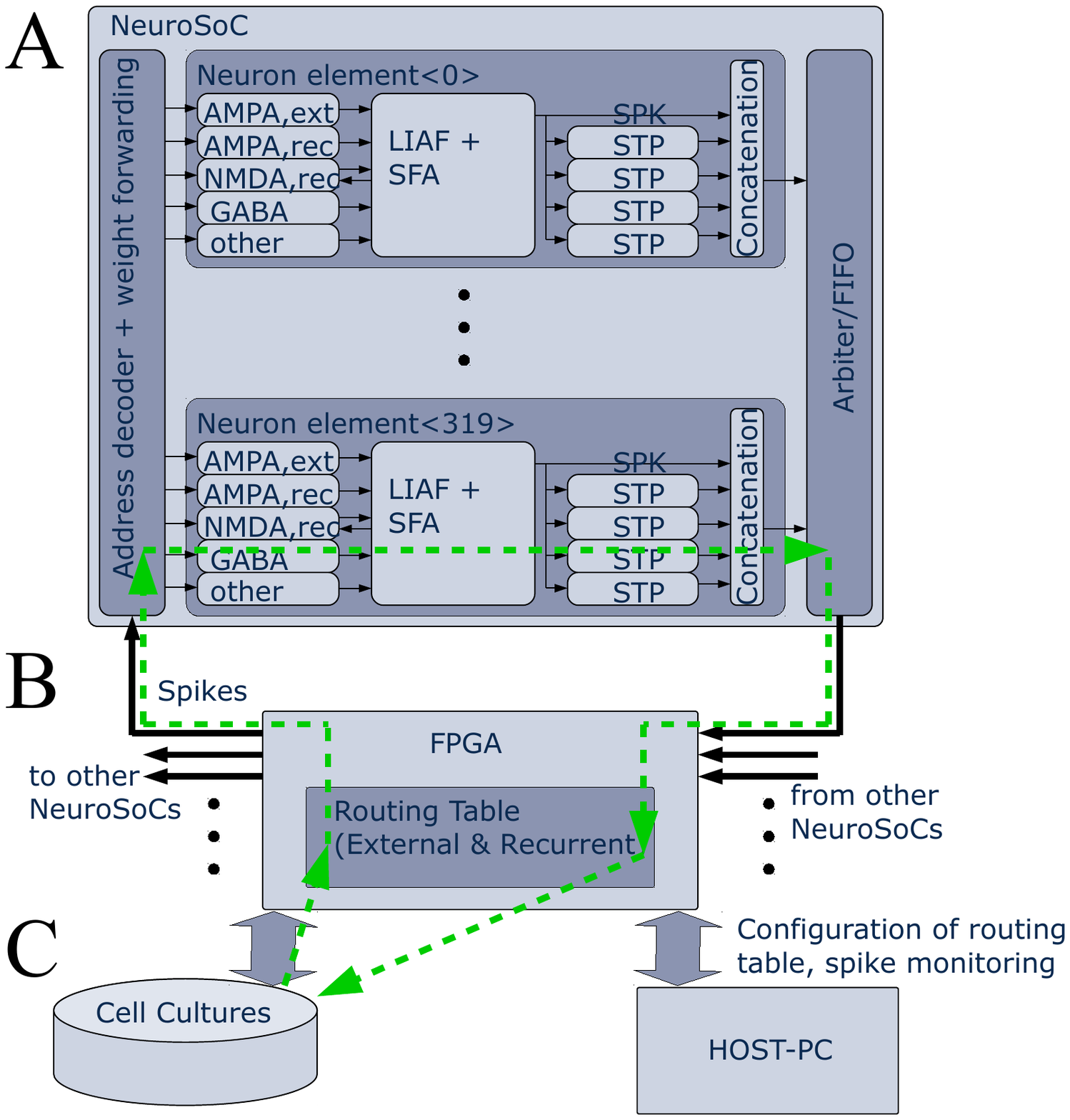}}
\end{center}
\vspace{-0em}
\label{fig_neurosoc_sys}
\end{figure}

\footnotesize{\textbf{Figure 2. Logical organization of the NeuroSoC system.} (\textbf{A}) Single NeuroSoC and its processing elements with sample spike processing chain illustrated in the green, dashed line. A spike enters via the address decoder and gets transmitted to the respective multi-synapse; this in turn generates a conductance change (excitatory or inhibitory) on the SC-neuron. Output spikes of the neuron influence its spike frequency adaptation (SFA) as well as its downstream presynaptic short-term plasticity (STP). The spike address plus its four STP values are concatenated into a pulse packet and sent off-chip. (\textbf{B}) The FPGA backplane contains playback memory for background stimulation and the routing table between the NeuroSoCs; as well, it enables communication with the neural culture setup and the controlling host PC. (\textbf{C}) Neural cultures are connected in a closed loop to the NeuroSoC system via Ethernet; the host PC enables configuration, live experiment monitoring and control.}
\normalsize
\vspace{+1cm}

The incoming synapses are conductance-based, with individually configurable characteristics, allowing to model AMPA, NMDA and GABA type synapses. One inhibitory conductance is triggered by the neuron itself and thus acts as spike-frequency-adaptation. The output of the neuron feeds into four digital short-term adaptation modules. Each of these can be configured individually and thus provide the neuron with different types of synaptic adaptation for its downstream connections.
All pulses are routed via an FPGA that also handles the routing in between the several NeuroSoC chips that make up the system. The NeuroSoC receives the incoming FPGA pulses via an address decoder, while an arbiter handles conflicts between outgoing pulses. The FPGA also handles the host communication and control. The Ethernet-based interface to the neural cultures is also implemented on the FPGA \cite{george2015event}. For simplicity, the clock for the SC circuits is supplied by the FPGA, but could in principle also be generated on chip \cite{eisenreich09}. 

Following the sample spike path from/to biology in Figure 2, for any spike entering the FPGA, a 6 bit synaptic weight and a target multi-synapse is read out of a routing table, allowing to set up arbitrary synaptic connections. The 6 bit fixed weight stored in the FPGA (Figure 2B) is supplemented by the 6 bit STP dynamic weight as output from the Markram/Tsodyks adaptation modules (Figure 2A).
For modeling the synaptic conductance in SC technique, a digital clock governs the switching between two capacitances (Figure 1C). A fixed frequency would result in a fixed conductance. Therefore, a digital circuit generates an exponentially-decaying frequency to emulate an instantaneous-onset, exponentially-decaying synaptic conductance. The absolute synaptic strength (essentially the highest frequency operating the switches at the onset of the conductance change) is governed by the multiplication of the two 6-bit weights described above. 
Each of the five synaptic conductances is implemented as a multi-synapse, meaning that multiple synaptic connections are modeled by overlaying them linearly on the same synaptic circuit. Hence the instantaneous increase will be re-triggered for each synaptic input on that conductance type and subsequent exponential decay performed on the accumulated value.

Figure 1C shows the analog circuits at the center of the neuron element.
In essence, this is a fully-differential SC implementation of a LIAF neuron with spike-frequency-adaptation and the conductance-based multi-synapses at the input. The NMDA synapses constitute a special case as they exhibit a non-linear dependence on the membrane voltage plus a saturation. We have approximated this by an analog computation along a piecewise linear curve \cite{noack14}. 

Due to its widely configurable behavior range, we employ the model of \cite{markram98} as presynaptic short term plasticity in the NeuroSoC. A fully digital translation of the model is used, as it was comparable in power and area to an SC implementation and offered superior programmability and repeatability. It transmits its 6-bit adaptation state off-chip for further routing in the FPGA. As mentioned earlier, this state is again used at the input of the next NeuroSoC as a dynamic synaptic weight value. In the same spirit as the multi-synapses, we did not implement a specific presynaptic adaptation circuit per single synapse configured in the system. Instead, four different realizations of the STP circuit driven by the same neuron provide for differently-configured flavors of presynaptic adaptation to postsynaptic neurons. That is, one STP channel could be configured for facilitation, one for depression, one for combined facilitation and depression \cite{noack11} and each postsynaptic neuron can be configured in the FPGA routing table 
to receive one of those flavors. 

Figure 3A shows the topology employed for the biohybrid experiment on the NeuroSoC system. At its center, all 2880 neurons of the system (distributed across the nine NeuroSoCs) are configured as an excitatory recurrent network with SFA. Recurrent connection probability is 0.007, i.e. on average a single neuron is recurrently connected to 20 other neurons. Feeding into this network is a background stimulus of 25 Poisson spike sources at 10 Hz, with each spike source densely connected to the recurrent network, with p=0.3 and a strong maximum synaptic conductance of g=20nS.
Also feeding into this network is the output of the spike detection operating on the 60 MEA channels, i.e. 60 channels of biological spikes. The SFA strength, strength of recurrent connections and the background are finely balanced for the network to experience short population spikes (i.e. behavior 2 of the range displayed in Figure 3B).
For more details on the theory-driven configuration strategy of the hardware network, see \cite{partzsch17}. The network can have spontaneous population spikes driven by the background stimulus as well as evoked spikes driven by the input from the neural cultures.
\vspace{+1cm}

\begin{figure}[H]
\vspace{-0em}
\begin{center}
\centerline{\includegraphics[trim= 0cm 6cm 0cm 2.5cm,width=160mm]{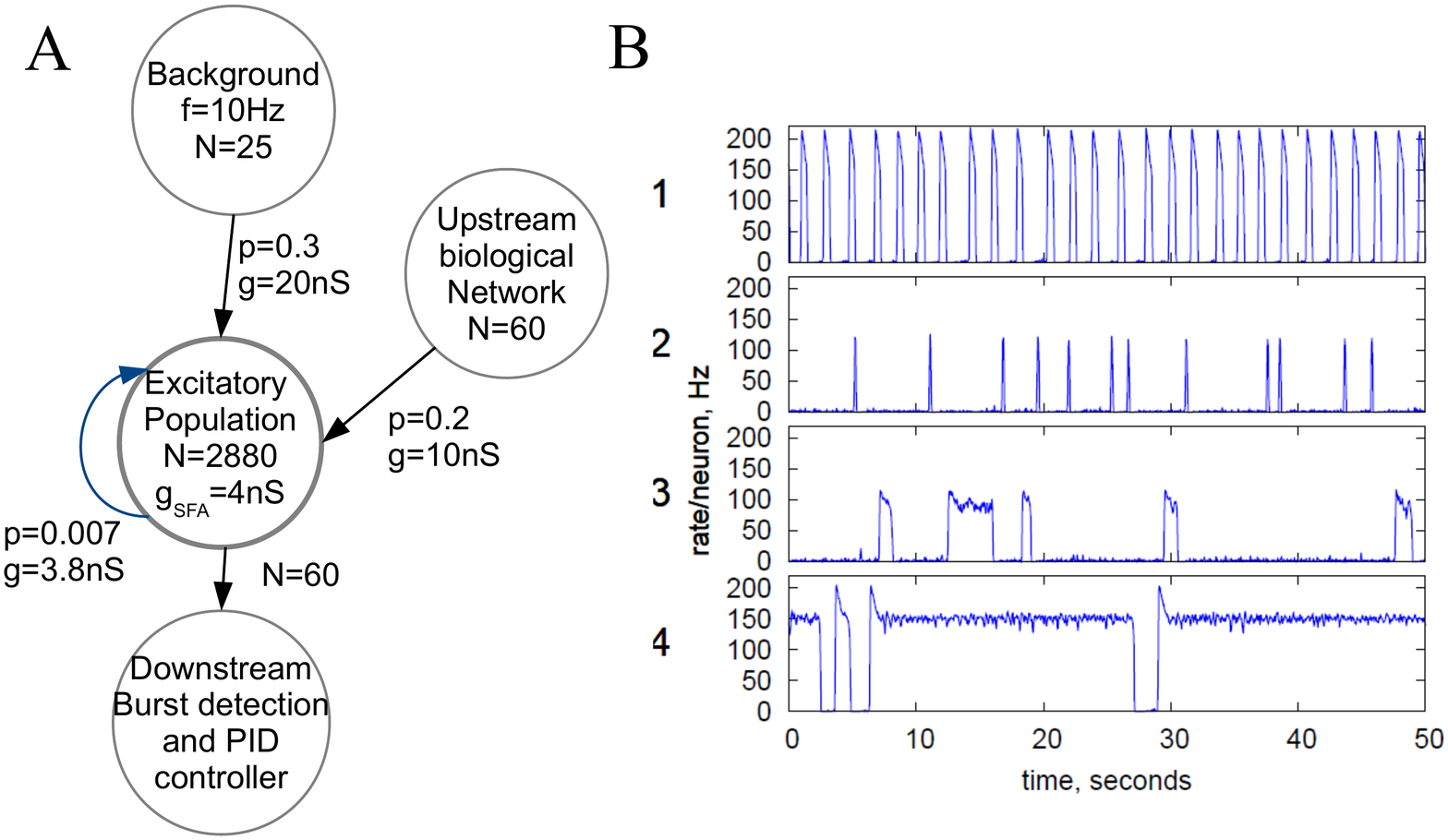}}
\end{center}
\vspace{-0em}
\vspace{-0em}
\label{fig_hybrid_setup}
\end{figure}

\footnotesize{\textbf{Figure 3. Configuration of the NeuroSoC system}. (\textbf{A}) Hardware network structure, with background stimulus provided by the FPGA and a single excitatory population configured with all 2880 neurons available; spike activity from the culture is added as additional synapses into a fraction of the excitatory network; the activity of 60 random neurons is transmitted back to the burst detection software and the subsequent PID controller. Each circle denotes a population of neurons/spike sources (e.g. the Poisson background provided by the FPGA at the top). N denotes the number of neurons/spike sources in said population, with the arrows indicating linkages and synapse direction; p is the connection probability between each pair of pre- and postsynaptic neuron of the two populations and g the conductance of the synapse. (\textbf{B}) Four sample population behaviors (population average firing rate, normalized to single neuron); the parameters of the excitatory network and the background stimulus can be configured for a range of behavior; for similarity with the neural culture behavior, behavior type 2 was chosen, resulting in the parameters given in A.}
\normalsize
\vspace{+1cm}

\textit{Cell preparation:} Cortical neurons were obtained from newborn rats (Sprague-Dawley) within 24 hours after birth using mechanical and enzymatic procedures described in earlier studies \cite{Marom2002d}. Rats were anesthetized by CO$_{2}$ inhalation according to protocols approved by the Technion's ethics committee. The neurons were plated directly onto a substrate-integrated multi electrode array and allowed to develop into functionally and structurally mature networks over a period of 2-3 weeks. The number of plated neurons was of the order of 450,000, covering an area of about 380 mm$^{2}$. The preparations were bathed in MEM supplemented with heat-inactivated horse serum (5\%), glutamine (0.5 mM), glucose (20 mM), and gentamycin (10 $\mu$g/ml), and maintained in an atmosphere of 37$^{\circ}$C, 5\% CO$_{2}$ and 95\% air, also during electrophysiological recording. 

\textit{Electrophysiology:} An array of Ti/Au extracellular electrodes, 30 $\mu$m in diameter, spaced 500 $\mu$m from each other and located in the center, was used (MultiChannelSystems, Reutlingen, Germany). A commercial amplifier (MEA-1060-inv-BC, MCS, Reutlingen, Germany) with frequency limits of 150-3,000 Hz and a gain of x1024 was obtaining data. Data was digitized using data acquisition board (PD2-MF-64-3M/12H, UEI, Walpole, MA, USA). Each channel was sampled at a frequency of 16 kHz. The insulation layer (silicon nitride) was pre-treated with polyethyleneimine (Sigma, 0.01\% in 0.1 M Borate buffer solution). Data acquired was analyzed using Matlab (Mathworks, Natick, MA, USA). 

\textit{Analyses:} 
Action potentials were detected by threshold crossing which is defined separately for each of the recording channels at the beginning of an experiment (6 $\times$ standard deviation of a 2 sec voltage trace). A refractory period of 6 msec was considered. Electrical activity detected often originates from several sources, typically 2-3 neurons, as each recording electrode is surrounded by several cell bodies. 
Detection of network bursts was performed on-line by threshold crossing of summed action potentials within 25 msec. Exact threshold was determined relative to 25\% of the active electrodes (typically the value of 20 action potentials). 

\textit{PI algorithm:}

\textit{ A Proportional-Integral-Derivative (PID) controller} was realized on the xPC target system \cite{Levine1996}. The input to the controller is the error signal,

\begin{equation}
e_{n}=P_{n}^{*}-\breve{P}_{n}\label{eq:01}
\end{equation}

where $P_{n}^{*}$ and $\breve{P}_{n}$ are the desired and estimated response probabilities at the $n^{th}$ stimulus, respectively. The output of the controller is generally composed of four components, 

\begin{equation}
A_{n}=A_{baseline}+g_{P}e_{n}+g_{I}\sum_{i=1}^{n}e_{i}+g_{D}(e_{n}-e_{n-1})\label{eq:02}
\end{equation} 

where $g_{P},g_{I}$ and $g_{D}$ are the proportional, integral and derivative gain parameters, respectively  (typically $g_{P}$ is 1 (400mV), $g_{I}$ is 0.2 (80mV) and $g_{D}$ is 0) and $A_{baseline}$ is the baseline amplitude bias (set according to an open stimulation amplitude required for reaching the desired reference response probability). In our systems possible stimulation amplitude ranges between 100mV-1150mV. Reaching the limits of stimulation amplitudes, results in a saturation of the input signal to a constant uncontrolled signal, while the integrative error value steadily increases. Hence, stimulation in the range of possible amplitudes, is required for maintaining control. Note that the controlling algorithm is reducing all the characteristics of responses to a stimulus, to a binary value representing whether a response occurred or not. 

\textit{On-line estimation of network response probability:} Let $s_{i}$ be an indicator function, so that $s_{i} = 1$ if the network generated a network burst within a predefined interval after the $i^{th}$ stimulus and $s_{i} = 0$ otherwise. 
We denote $\tilde{P}_{n}$ the estimation calculated at time $t > tn$, based on the set of responses $\{s_{1}, s_{2},..., s_{n}\}$ to stimuli given at times $\{t_{1}, t_{2},..., t_{n}\}$. A weighted average was realized by using the recursive formula,

\begin{equation}
\tilde{P}_{n}=(1-e^{-\frac{t_{n}-t_{n-1}}{\tau}})\cdot{s_{n}}+e^{-\frac{t_{n}-t_{n-1}}{\tau}}\cdot{\tilde{P}_{n-1}}\label{eq:03}
\end{equation}

$\tau$, the estimation time-constant, was typically set to 250 sec. In order to reduce the chances of identifying a `spontaneous' network burst (i.e. one that was not elicited by stimulation) as an evoked one, the time window of response detection was set to 10-800 msec.

\section{Results}


The neuromorphic chip takes an approach of system-on-chip integration, i.e. we implement common peripheral components such as voltage biases directly on the chip together with the neuromorphic components. In keeping with conventional semiconductor chip naming, we call our chip NeuroSoC (Neuromorphic System-on-Chip, see  Figure 1B). Nine of the NeuroSoCs are grouped in a support system which encompasses a Field-Programmable Gate Array (FPGA). The FPGA sets up the synaptic connectivity between the neurons on the individual NeuroSoCs. Also, it serves as interface to the \textit{in-vitro} neural culture Micro-Electrode Array (MEA) and as connection to the configuration and control PC system (see Figure 1A). Overall, the system in Figure 1A contains 2880 neurons, 14400 conductance-based multi-synapses (equivalent to 12.7M synapses) and 11520 presynaptic short term plasticity circuits. 

As mentioned in the introduction, the neurons and synapses on the NeuroSoC are carried out in the SC circuit technique. This means that conductances are modelled by capacitances connected by switches, where the switching frequency in connection with the capacitance determines the conductance value (see Figure 1C). A specialized switching circuit has been used to reach the necessary timescales in the order of 1 second \cite{mayr14c}, which are otherwise hard to obtain in semiconductor circuits due to leakage currents \cite{roy03}. In SC circuits, time constants and gain parameters depend on capacitance ratios and switching frequencies and not on process-dependent transistor parameters, thus the NeuroSoC exhibits significantly more robust and reproducible behavior than conventional subthreshold neuromorphic circuits. Previously, SC circuits were used for simple membrane leakage current generation and synaptic transmission \cite{folowosele09b,vogelstein07}, however, here we implement more involved, 
biologically realistic models, with multiple individually configurable conductance-based synapse types and spike-frequency adaptation \cite{noack14,noack14b}. Aided by the good statistical adherence of our implementation to the model behavior, we can use a theory-guided approach \cite{giulioni12,partzsch17} to configure a recurrent network for a wide range of dynamical regimes. Figure 3A illustrates the network configured on all 2880 neurons of the NeuroSoC system. Specifically, we can obtain bursting behavior as observed in cultured networks \cite{Keren2014} (see Figure 3B for obtainable behavior range). We use the bursting behavior seen in graph 2 of Figure 3B as viable counterpart to the in-vitro network.

The feasibility of integrating the hardware network with the biological culture is demonstrated using a control circuit, which considers the coupled biohybrid activity as of a single unified network. The upstream culture is directly stimulated, while the monitored output is of the hardware network. The two parts are connected in real time by sending all recorded culture activities as inputs to the hardware network in iterations of 0.5 msec. The hardware network activity is recorded from 60 randomly sampled nodes, the activity of which is sent back to a PI controller - for configuration of the next stimulation amplitude (the transmission of data packets from culture to hardware and back, spanned in the order of 1 msec). The consequent stimulation amplitude, is then again delivered to the culture network (as illustrated in Figure 4A).
The control in this case is of \textit{network response probability}, which is the probability of evoked bursts occurrence following a stimulation. The control target values of response probability is a sin wave function (inset of Figure 4B). Both networks were congruently following this target (see \cite{Wallach2011} and \cite{Keren2014} for further technical details of the control algorithm). Under this setting, hardware activity, otherwise spontaneously active in a fluctuating manner, is highly aligned to the upstream culture activity (exemplified in Figure 4C). 
\vspace{+1cm}

\begin{figure}[H]
\vspace{-0em}
\begin{center}
\centerline{\includegraphics[trim= 0cm 3cm 6cm 0cm,width=160mm]{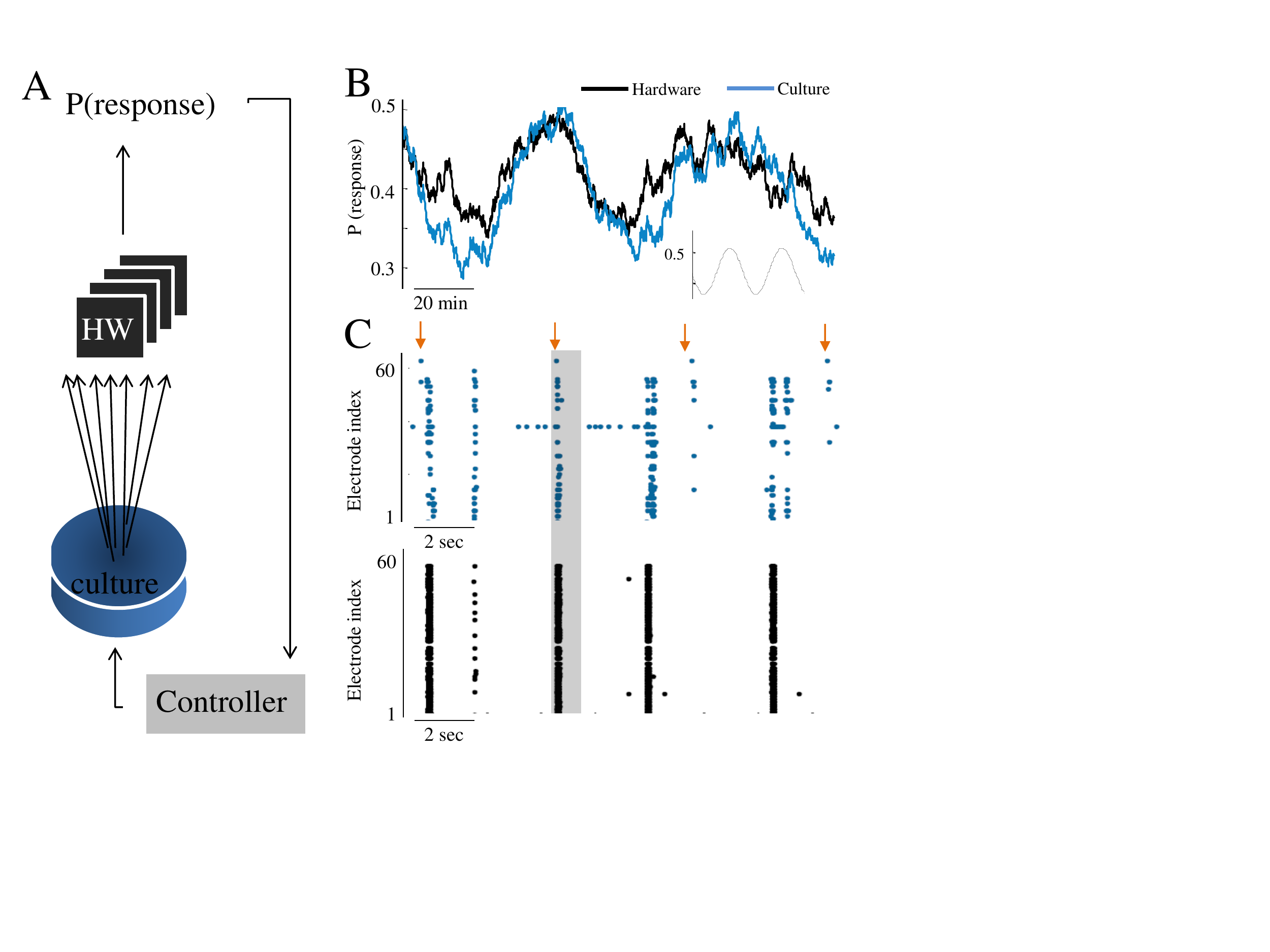}}
\end{center}
\vspace{-0em}
\vspace{-0em}
\label{fig_hybrid_two_levels}
\end{figure}

\footnotesize{\textbf{Figure 4. Hybrid indirect closed-loop control}. (\textbf{A}) Scheme of the coupling between the culture and neuromorphic device: the stimulation is applied to the culture, and the PI controller calculates the appropriate stimulation amplitude by the activity of the downstream hardware. (\textbf{B}) The output of response probability of culture and hardware under control (see methods for the control algorithm). (\textbf{C}) Extracts of aligned activity as recorded in culture (blue) and hardware (black). Each point depicts a single spike detected in one of the electrodes (indexed in the vertical axis). Arrows depict stimulation times and shaded regions depict the time window for detection of an evoked burst (10-800 msec after stimulation).}
\normalsize
\vspace{+1cm}

\newpage
To explore a possible reflection of controlled activity features in the downstream module when only a single upstream module is controlled, the hardware is coupled to the activity trace of a culture being controlled independently to two distinguished response rates. We denote the two different levels of network evoked responses \textit{high} for 0.7 response probability and \textit{low} for response probability of 0.2-0.3. Each mode is maintained for more than an hour. As shown by three different experiments, both culture and hardware response traces, are congruently high or low (Figure 5A, depicted blue and black respectively). In this setting, culture response dynamics are dominating the hardware's activation (as hardware bursts are predominantly preceded by a culture burst). As demonstrated in Figure 5B, the stimulation amplitude values required for maintaining this controlled activity in the culture, are higher or lower respectively to the intended mode of activity. In contrary, the culture evoked burst amplitudes - the stimulating source of the hardware network - are not modified accordingly to the response level of the culture (Figure 5C). Hence the modified stimulation amplitude is only affecting directly the upstream culture (by modifying response probability), while the downstream part of the network is controlled by the occurrence frequency of culture responses (rather than by the intensity upstream activity). 
\vspace{+1cm}

\begin{figure}[H]
\vspace{-0em}
\begin{center}
\centerline{\includegraphics[trim= 0cm 1cm 15cm 0cm,width=90mm]{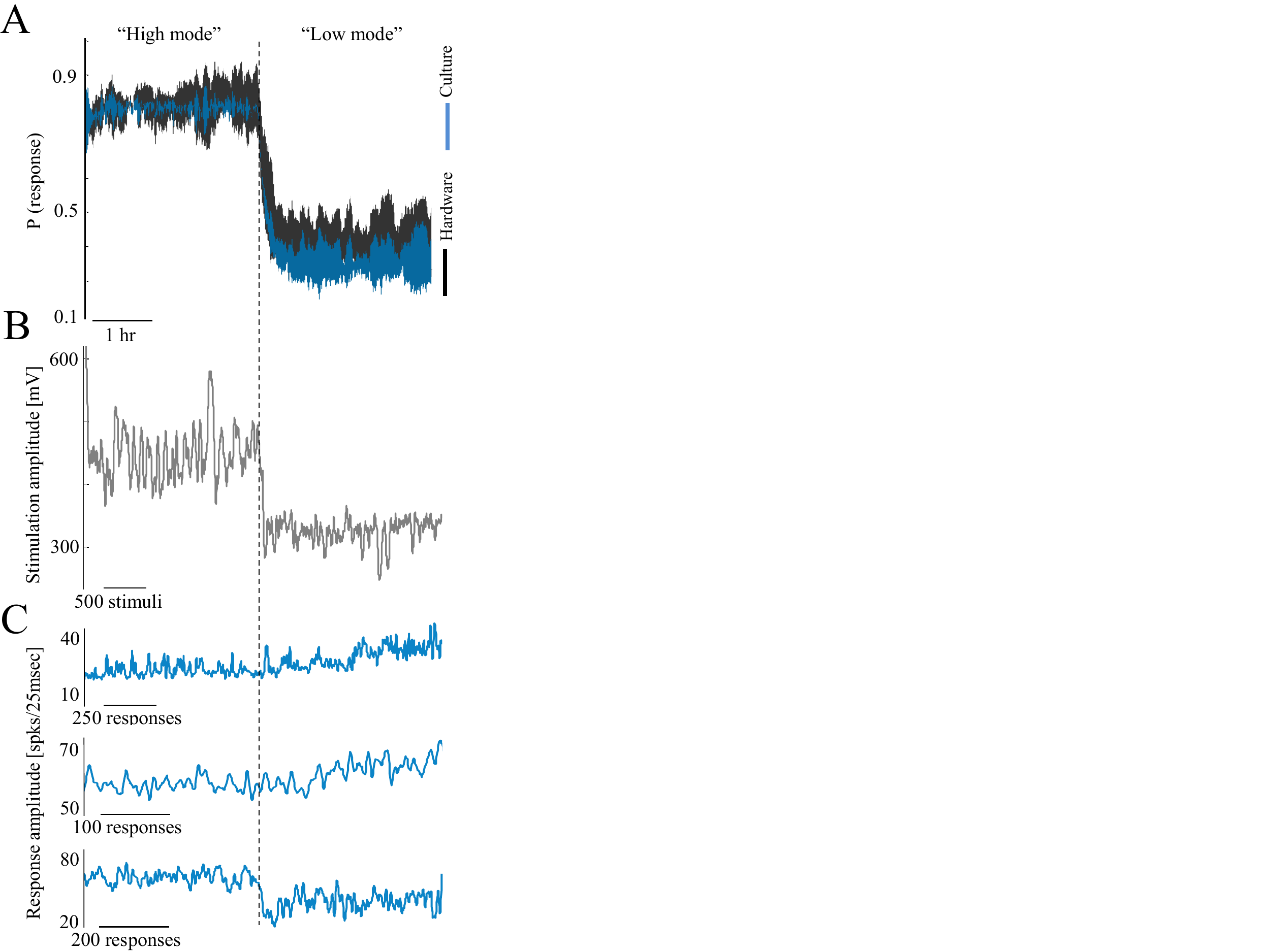}}
\end{center}
\vspace{-0em}
\vspace{-0em}
\label{fig_hybrid_upstream_downstream}
\end{figure}

\footnotesize{\textbf{Figure 5. Control to two response probability levels}. (\textbf{A}) Cultures controlled to a high response probability for 3 hours, followed by control to low response level (blue). Trace margins are std values across two cultures controlled similarly. Hardware response probability, when fed as inputs with this culture activity trace, is depicted similarly in black. (\textbf{B}) The respective stimulation amplitudes required for controlling one of the cultures, as computed by the control algorithm. Note that higher stimulation amplitudes are required for higher response probability (all three experiments show a similar trend). (\textbf{C}) Evoked burst sizes of three cultures, showing no consistent relation to the response probability level.}
\normalsize
\vspace{+1cm}

Moreover, we demonstrate that by externally enforcing networks response probability to different levels, other activity features are changing respectively. Propagation delays become faster during a high response rate and slower during a low response state. Propagation delays in this context are defined as the inter-spike-intervals during the recruitment rate of evoked bursts. Figure 6A demonstrates that this effect on propagation times, occurs congruently in both culture and hardware. Response latency values are as well altered respectively, being shorter in higher response mode, congruently with previous findings of stimulation amplitude-latency relations \cite{Keren2014}. This impact of control affects similarly both upstream culture and the downstream hardware network (see inset to 6A). We also find that the duration of evoked bursts is altered between the high and low response modes, in a similar direction in both upstream and downstream modules. From a spatial perspective, maintaining the response rate at a higher or lower level, results in an exploitation of different propagation paths across the culture (Figure 6B, left, a comparison of the distance metric between all response recruitment orders, note two distinguishable groups of responses). We find this to be slightly reflected as well in the downstream hardware module (Figure 6B, right panel), where it is evident that during the higher response mode more similar propagation paths are used. 
This suggests that propagation paths in the culture could be dictated by the value of stimulation amplitude, which on the other hand is not reflected as much in the downstream hardware module.
\vspace{+1cm}

\begin{figure}[H]
\vspace{-0em}
\begin{center}
\centerline{\includegraphics[trim= 0cm 6cm 12cm 0cm,width=110mm]{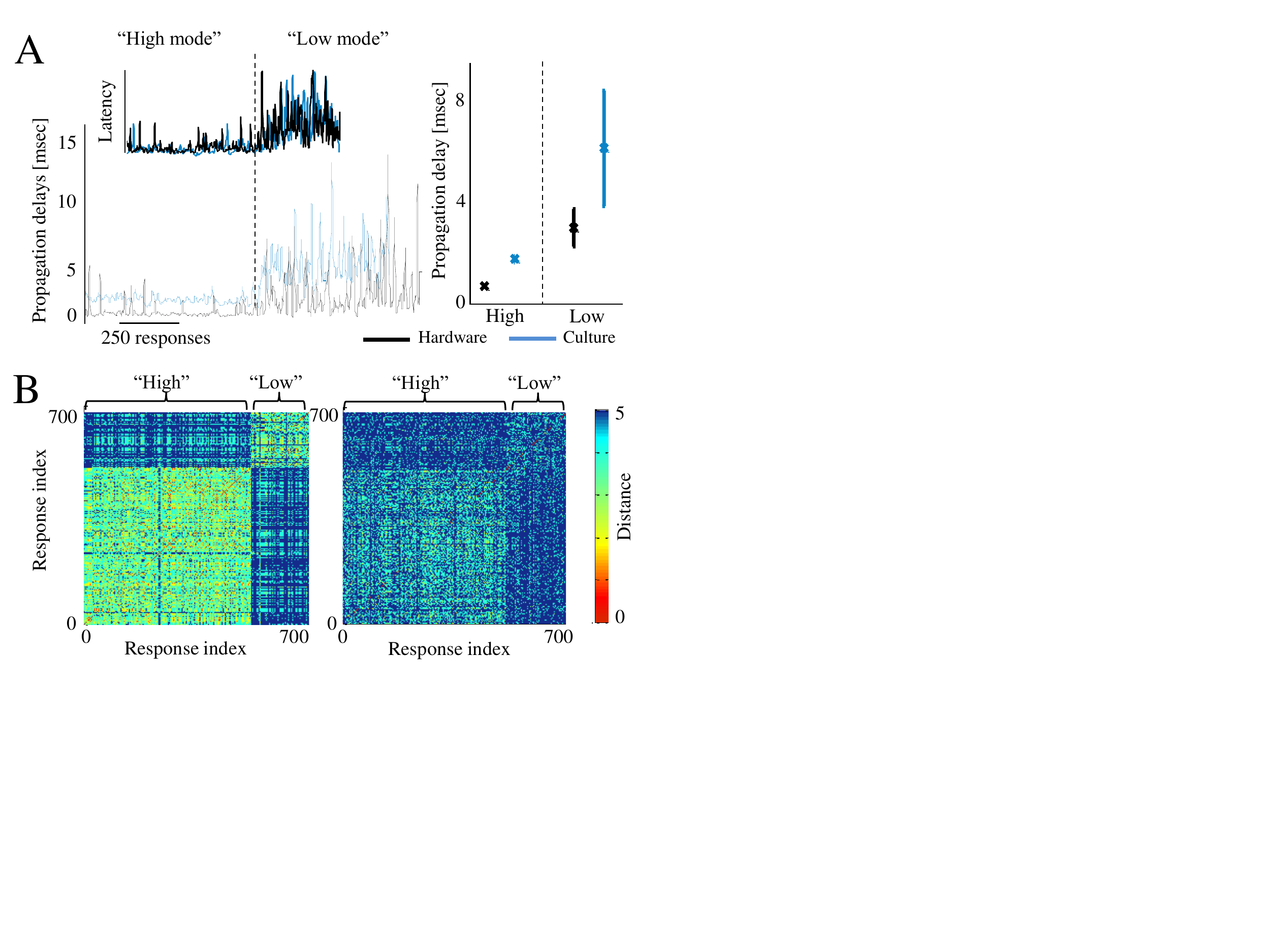}}
\end{center}
\vspace{-0em}
\vspace{-0em}
\label{fig_hybrid_upstream_downstream}
\end{figure}

\footnotesize{\textbf{Figure 6. Upstream and downstream impacts of control}. (\textbf{A}) Delays between spikes during recruitment phase of evoked responses, under control to either high or low response rate, in a single coupled experiment (blue depicts the upstream culture and black the downstream hardware). Inset shows the respective alteration of latency values, calculated as the time from stimulation to the burst peak. Right panel: average values across three experiments, error bars show std (in respective colors). (\textbf{B}) Bursts recruitment orders of culture (left) and hardware (right), during control to either high or low response mode, in a single coupled experiment. Pair-wise difference between propagation paths are presented (the Levenshtein distance metric), for five first electrodes to respond, across all evoked bursts consecutively (see \cite{Shahaf2008} for details). The distance between order vectors of evoked burst $i$ and $j$ is color coded in pixel ($i$,$j$) between red (0, identical) and blue (5, no identity).}
\normalsize
\vspace{+1cm}

As denoted in Figure 3A, the connectivity from culture to hardware is a random selection of culture units per hardware neuron ($20\%$), while the 60 neurons monitored in the hardware are randomly selected among the 2880 neurons. Hence, any spatial aspects of upstream activity are not directly transferred to the downstream module by the virtue of coupling (all results were replicated also with a reduced coupling connectivity strength of 10\% instead of 20\%).
However, the downstream hardware seem to still be affected to some extent by the temporal order of spikes, as demonstrated by the results in Figure 6B.

%

\section{Concluding remarks}

By integrating a neural network \textit{in-vitro} with a neuromorphic hardware network, we were able to study impacts of control on a modular biohybrid. In this modular organization the inputs to the downstream network can be measured and modified. We demonstrate the feasibility to control indirectly this unified network - by stimulating one module according to the output of the second module. This model of such a control circuit reflects a reasonable physiological setting, aiming at controlling downstream symptoms generated by upstream neural irregularities \cite{Walker2012}. 
The downstream hardware follows upstream culture dynamics, also when the culture module is the only part being directly stimulated.
This was feasible even when using sparse connectivity from culture to hardware (as low as $10\%$ connection probability), and sampling only a subset of the hardware activity (a subset of 60 out of 2880 neurons, similar to the electrodes-culture recorded region relation).
    
Open and closed-loop stimulation regimes have been shown previously to induce alterations of neural functional properties, in the region of control \cite{jimbo1998,jimbo1999, Madhavan2006, Chao2007, Chiappalone2008, Vajda2008, Bologna2010, Feber2010, Keren2014}. We explored control induced alterations in the downstream module, when controlling only one module of the coupled construct with a closed-loop stimulation. We show that controlling the upstream culture part of the two coupled networks, generated similar response dynamics also downstream. Possible microscopic level influences were also demonstrated on both modules, mainly by faster propagation of evoked bursts due to control to high response probability. This potentially reflects increased excitatory resources during control to higher response rates \cite{Haroush2015}, occurring efficiently on both parts. Faster propagation during recruitment, is also congruent with the relation found between stimulation amplitude and response latency- where higher stimulation amplitudes evoke shorter latencies of network bursts \cite{Keren2014}, as also evident in this study. During control to a state closer to spontaneous activity rates (low response probability), propagation delays are slower and the overall response latencies are longer. The alterations of network activity are evident on both modules, even though the downstream module is not directly stimulated - this suggests that such changes can be induced also by higher burst rates, in addition to a direct stimulation with high amplitudes. 
These results demonstrate how local temporal changes induced by external control, can result in a more global functional impact. 
Closed-loop control - a powerful external intervention - appears to modify efficiently neural activity, and moreover also at downstream regions. This is a significant consideration when developing brain-neuromorphic interfaces and external control algorithms of modular networks.

\newpage
\renewcommand\refname{References}
\bibliographystyle{plos2015}
\bibliography{RefHWpaper,ref}
\makeatletter
\renewcommand{\@biblabel}[1]{\quad#1.}
\makeatother
\end{document}